\documentclass[12pt,preprint]{emulateapj}

\usepackage{url}
\usepackage{bm}
\usepackage{color}
\usepackage{pdflscape}

\begin{document}

\title{LkCa 15: A Young Exoplanet Caught at Formation?} 
\author
{Adam L. Kraus\altaffilmark{1},
Michael J. Ireland\altaffilmark{2,3,4}
}

\altaffiltext{1}{Hubble Fellow; Institute for Astronomy, University of Hawaii at Manoa, 2680 Woodlawn Dr, Honolulu, HI 96816, USA}
\altaffiltext{2}{Department of Physics and Astronomy, Macquarie University, NSW 2109, Australia}
\altaffiltext{3}{Australian Astronomical Observatory, PO Box 296, Epping, NSW 1710, Australia}
\altaffiltext{4}{Sydney Institute for Astronomy (SIfA), School of Physics, NSW 2006, Australia}

\begin{abstract}

Young and directly imaged exoplanets offer critical tests of planet-formation models that are not matched by RV surveys of mature stars. These targets have been extremely elusive to date, with no exoplanets younger than 10--20 Myr and only a handful of direct-imaged exoplanets at all ages. We report the direct imaging discovery of a likely (proto)planet around the young ($\sim$2 Myr) solar analog LkCa 15, located inside a known gap in the protoplanetary disk (a ``transitional disk''). Our observations use non-redundant aperture masking interferometry at 3 epochs to reveal a faint and relatively blue point source ($M_{K'}=9.1 \pm 0.2$, $K'-L'=0.98 \pm 0.22$), flanked by approximately co-orbital emission that is red and resolved into at least two sources ($M_{L'}=7.5\pm0.2$, $K'-L'=2.7\pm0.3$; $M_{L'}=7.4\pm0.2$, $K'-L'=1.94\pm0.16$). We propose that the most likely geometry consists of a newly-formed (proto)planet that is surrounded by dusty material. The nominal estimated mass is $\sim$6 $M_{Jup}$ according to the 1 Myr hot-start models. However, we argue based on its luminosity, color, and the presence of circumplanetary material that the planet has likely been caught at its epoch of assembly, and hence this mass is an upper limit due to its extreme youth and flux contributed by accretion. The projected separations ($71.9 \pm 1.6$ mas, $100.7 \pm 1.9$ mas, and $88.2 \pm 1.8$ mas) and deprojected orbital radii (16, 21, and 19 AU) correspond to the center of the disk gap, but are too close to the primary star for a circular orbit to account for the observed inner edge of the outer disk, so an alternate explanation (i.e., additional planets or an eccentric orbit) is likely required. This discovery is the first direct evidence that at least some transitional disks do indeed host newly-formed (or forming) exoplanetary systems, and the observed properties provide crucial insight into the gas giant formation process.

 \end{abstract}

\section{Introduction}

In the past 15 years, indirect searches for extrasolar planets (i.e., radial velocity and transit surveys) have discovered over 500 confirmed planetary companions to other stars \citep{Wright:2010lr}, spurring explosive growth in the field of comparative exoplanetology. However, virtually all of these planets orbit around old  stars ($\tau \ge 1$ Gyr), typically at orbital radii smaller than that of Jupiter. Direct detection via high-resolution imaging holds great promise for extending comparative exoplanetology across the full range of planetary ages and orbital radii. Direct detection programs are vital for studying the outer regions of extrasolar systems (where analogs to our own gas giant planets should reside) since those planets are inaccessible to transit searches and can only be discovered with decades-long surveys by RV programs. Direct detection also offers a window into the detailed atmospheric and evolutionary properties of exoplanets since they are amenable to photometric and spectroscopic study. Finally, young exoplanets should be much hotter and more luminous than their older counterparts \citep{Marley:2007qx,Fortney:2008lo}, so they are easier to detect and should offer a window into the very early stages of planet formation and evolution. Direct-imaging techniques still face significant technical challenges, but those challenges are being overcome with innovative new observational techniques that have begun to yield the first discoveries \citep{Marois:2008zt,Kalas:2008jl,Lagrange:2009fc}.

Young planets and outer planets will provide critical new tests of the formation and evolution of planetary systems, as the two competing models of planet formation (``core accretion'' \citep{Pollack:1996dk} and ``disk instability'' \citep{Boss:2001sd}) make very different claims about when and where planets should form. Disk instability should preferentially form planets in outer solar systems, and typically will do so within $\sim$1 Myr after a star has formed. In contrast, core accretion is much more efficient in forming planets with small orbital radii, and the assembly process should require $\sim$3--5 Myr to form gas giants analogous to our own Jupiter and Saturn. Furthermore, almost all gas giant planets are expected to migrate both inward and outward in their orbits due to gravitational interactions with the protoplanetary disk and with smaller rocky bodies \citep{Ida:2004vh,Tsiganis:2005fk}. Almost all of the gas giant planets around solar-type stars with orbital radii of $\le$2--3 AU are thought to have formed at larger radii and migrated inward, so their frequency and orbital properties are not necessarily representative of most systems. The most demanding tests of planet formation models can come only from direct observations of young planetary systems as they form.

We have learned much about planet formation from other advances over the past decade. Mid- and far-infrared observations with the Spitzer Space Telescope have cast new light on the formation, evolution, and lifetimes of the circumstellar disks where planets form \citep{Hillenbrand:2008pd}. Similarly, new observations by mm/submm observatories have established the masses and sizes of disks, especially on spatial scales of 10--100 AU where dust is too cool to emit in the mid-IR or far-IR \citep{Andrews:2005qf,Andrews:2011lr}. In both cases, observations have also discovered a rare, intriguing class of objects: protoplanetary disks where the mid-IR spectral energy distributions or resolved images (from submm/mm wavelengths) reveal gaps or inner holes \citep{Calvet:2005xf,Espaillat:2007rq,Brown:2009sh}. These gaps could be cleared by the gravitational influence of other bodies \citep{Ireland:2008kx,Huelamo:2011fk}, such as stellar binaries or gas giant planets. In cases where binary companions are ruled out, then these gaps serve as signposts of likely ongoing planet formation. These likely sites of planet formation are natural targets for pushing the boundaries of observational capabilities. Giant planets should be present, and the geometry of the disk gaps can even demonstrate their locations. To this end, we have begun a survey to directly image exoplanets inside the gaps of protoplanetary disks using a recently developed technique, nonredundant mask interferometry (NRM). 

One of the first targets we chose for our survey was the young \citep[$2^{+2}_{-1}$ Myr; ][]{Kraus:2009fk} solar analog LkCa 15, which is located in the Taurus-Auriga star-forming region and is known to have a massive (55 $M_{Jup}$) circumstellar disk \citep{Andrews:2005qf}. Detailed modeling of the disk's mid-infrared spectrum has demonstrated that it appears to have a sizeable gap \citep{Espaillat:2007rq}. LkCa 15 shows near-infrared emission from warm dust in the inner $<$1 AU, plus mid- and far-infrared emission from cold dust outside $>$50 AU. However, there is a deficit of emission at 10--20 $\mu$m, indicating the presence of a gap at intermediate radii that is cleared of dust. Subsequent observations at longer wavelengths that directly trace the dust also demonstrate a deep paucity of emission at separations of $<$55 AU \citep{Andrews:2011lr}, and near-infrared imaging may have observed reflected light from the inner edge of the disk \citep{Thalmann:2010mz}. We previously observed LkCa 15 with NRM in the $K'$ band to determine if this cleared region could indicate the presence of a binary companion \citep{Kraus:2011qy}, but found no companions with a mass of $> 12 M_{Jup}$). We therefore made the system a high priority for our planet-search program, which is obtaining significantly longer observations than our previous study ($\ge$4 hours, versus $\sim$20 minutes).

In this paper, we report the discovery of a likely (proto)planet orbiting LkCa 15, which we have caught at the epoch of formation. In Section 2, we describe our observations of LkCa 15 and the methods we used to analyze our data. In Section 3, we describe our discovery of the apparent (proto)planetary companion and circumplanetary material, its apparent properties, and the alternative explanations for our observations that we are able to rule out. Finally, in Section 4, we discuss some of the implications of our discovery for planet formation and for the nature of transitional disks.

\section{Observations and Data Analysis}

\subsection{Nonredundant Mask Interferometry of LkCa 15}

The sensitivity of adaptive optics imaging for close companions ($<$2--3$\lambda$$/D$) is limited by imperfect calibration of the primary star's point spread function (PSF); the PSF width and shape change under different atmospheric conditions, and quasistatic image artifacts (``speckles'', which resemble faint companions) are superimposed on the image by uncorrected atmospheric turbulence and by optical imperfections in the telescope itself. The technique of non-redundant aperture masking (NRM) has been well-established as a means of achieving the full diffraction limit of a single telescope \citep{Nakajima:1989zl,Tuthill:2000ge,Tuthill:2006kq}. NRM uses a pupil-plane mask to block most of the light from a target, resampling the primary mirror into set of smaller subapertures that form a sparse interferometric array. Rather than an image of the target, the science camera then observes its interferometric fringes. NRM allows for superior calibration of the stellar primary's point spread function and elimination of speckle noise by the application of interferometric analysis techniques. In particular, the measurement of closure phases allows for strehl-independent calibration of the stellar PSF and the cancelling of low-order phase errors that cause speckle noise in conventional imaging. NRM observations can yield contrasts of $\Delta K \sim$6 mag at $\lambda /D$ and $\Delta K \sim 4$ mag at 1/3 $\lambda /D$ even in very short observations. Some of the unique results of past high-contrast NRM observations include a measurement of one of the first dynamical masses for a brown dwarf, GJ 802 B \citep{Ireland:2008yq}, several studies of the multiplicity of young stars \citep{Ireland:2008yq,Ireland:2008kx}, and the discovery of a potentially substellar candidate companion to another transitional disk host, T Cha \citep{Huelamo:2011fk}.

We observed LkCa 15 over the course of three observing runs using the Keck-II 10m telescope in November 2009, August 2010, and November 2010. All observations were conducted with the facility AO imager, NIRC2, which has aperture masks installed in the cold filter wheel near the pupil stop. All of our observations used a 9-hole aperture mask, which passes 11\% of the total incident flux through nine 1.1m subapertures that span baselines of 1.5--9.2m. This choice maximizes the throughput, as the other option (an 18-hole mask) passes half as much incident flux and can only be used with narrowband filters (due to wavelength-dependent dispersion in broadband filters) that are $\sim$10\% as wide as the corresponding broadband filters. The nine subapertures yield 28 independent baseline triangles about which closure phases are measured. All NRM observations operate in a subarray mode of the narrow camera, which has a pixel scale of 9.963 ms/pix , and we conducted our observations using the $L'$ (3.43-4.13 $\mu$m) and $K'$ (1.96-2.29 $\mu$m) broadband filters. The observations spanned most of several nights, so LkCa 15 was observed at airmasses ranging from 1.0 to 1.9. We summarize the average seeing on each night in Table 1.

Each observing sequence consisted of multiple ``visits'' of LkCa 15, alternating with observations of independent calibrator stars. These calibrators are essential point-spread-function calibrators, which for aperture-masking observations are used to estimate the systematic non-zero closure-phases (e.g., third order effects of phase aberrations within each sub-aperture). We chose these calibrators to be near LkCa 15 on the sky ($<$7$^o$ separation) and to have similar brightness in the optical (for similar adaptive optics performance) and to be brighter in the near-infrared (so that the Poisson noise would be small compared to that from LkCa 15). The number of observations taken at each epoch and the calibrators used are given in Table 1. All of these stars were chosen from our previous binary survey \citep{Kraus:2011qy} and are known to have no stellar companions at angular separations of $>$20 mas and no brown dwarf companions at $>$40 mas. Each individual calibrator is observed much less often than LkCa 15, so any calibration error from faint companions to the calibrators should be negligible.

Each ``visit'' consisted of a long sequence of exposures. In the $L'$ observations, we obtained 20 exposures of 20s each, yielding a total integration time of 400s. Data were taken in a two-point dither mode with the interferogram in opposite quadrants of the 512x512 subarray of the NIRC2 camera. We used this subarray mode to minimize readout overhead. In the $K'$ observations, we obtained 12 exposures of 20s per visit because we expected calibration error to be more significant at $K'$ than at $L'$ (due to the lower strehls delivered by the AO system), and thus we wanted the science and calibration observations to be as simultaneous as possible. We also did not dither, since the thermal backgrounds at $K'$ are negligible given the brightness of our targets (unlike for $L'$) and therefore sky subtraction was unnecessary. We summarize the history of all ``visits'' in Table 1.

\subsection{Data Analysis}

The data analysis up to the calibration step was identical to that used in previous papers \citep{Ireland:2008yq,Kraus:2008zr}. However, most of the subsequent steps were developed or refined when we discovered that our initial fits (for a single point-source companion) had high residuals, indicating the possibility of more complex structure. To briefly summarize the identical initial steps, the images were flat-fielded and bad pixels were removed by interpolating between neighbouring pixels. The image was then multiplied by a super-Gaussian window function of the form $\exp(-ar^4)$, with $r$ the radius in pixels from the center of mass of the interferogram. A two-dimensional Fourier transform was then made of each exposure in a visit, and this Fourier transform was point-sampled at the positions corresponding to the baseline vectors in the aperture mask. For visit $k$ we then computed the vector of mean uncalibrated closure-phases $\bm{x_k}$ and the standard error of the mean  $\bm{\sigma_k(x_k)}$.

In the past, we have generally calibrated the closure phase simply by subtracting off the uncalibrated closure-phases measured for calibrator stars observed closest in time. We initially used this method for analysis in this paper, detecting the same key structures as reported below. An independent analysis of the 2009 epoch using the SAMP code \citep[][]{Lacour:2011kx} also produced consistent results (S. Lacour, priv. comm). However, deciding on which calibrators are ``closest'' in time and how many to use is a somewhat arbitrary process, and in addtion, we must consider variable atmospheric dispersion in the L-band \citep[as discussed in ][]{Hinkley:2011lr}. 

For results reported here, we improved our previously used technique by choosing an optimal linear combination of calibrators for each target star visit. The detailed motivation for this algorithm will be described in Ireland (2011, in prep), but we describe the essential components of the algorithm here. Firstly, rather than using correlated closure-phases as the primary observable in the fitting algorithm, we used statistically independent linear combinations of closure-phases $\bm{x_k}$. This is a similar approach to that used by \citet[][]{Martinache:2010tj} for his kernel phases, however we guarantee statistical independence using the measured closure-phase covariance matrix rather than assuming all Fourier phases to be independent. Our optimal linear combination of calibrators is such that the sum of the mean-square residuals of calibrated observables $\bm{x_t} - \Sigma_{k=1}^{N_c} a_k \bm{x_k}$ and the estimated variance in our observables is minimised. The sum is given by:

\begin{equation}
 S = |\frac{\bm{x_t}+ \Sigma_{k=1}^{N_c} a_k
  \bm{x_k}}{\bm{\sigma^2(x_t)} + \Delta_t^2}|^2 +
 |\frac{\Sigma_{k=1}^{N_c} a_k^2(\bm{\sigma_k^2(x_k)} + \Delta_t^2)}
    {\bm{\sigma^2(x_t)} + \Delta_t^2}|^2.
\label{eqnCRAZY}
\end{equation}

Here $N_c$ is the number of calibrators observed on a night, $\bm{x_t}$ is the vector of target closure-phase combinations being calibrated, the calibrator weights are $a_k$ and the systematic error component is $\Delta_t$. The systematic error component  $\Delta_t$ is set so that the reduced chi-squared for each target visit is no more than unity. Like the LOCI algorithm used to reduce AO direct imaging data \citep{Lafreniere:2007qy}, finding an optimal linear combination of calibrators in this way tends to reduce the significance of a real companion in the data. For this reason, for the fits consisting of several point-sources we used this technique to determine if any given model was significant, then re-computed the calibration optimization using Equation~\ref{eqnCRAZY} where the target phase combinations $\bm{x_t}$ had the phases of the best fit global model subtracted (with 3 additional companions: see below). The difference this updated calibration caused to the final fitted parameters was less than 1.5\,$\sigma$ in all cases. 

Typical values of $\Delta_t$ required to achieve unity reduced chi-squared were comparable to or larger than the $\sim$0.8 degrees closure-phase scatter in the L' filter when the target phase combinations $\bm{x_t}$ did not include the best fit model, but were zero $\sim$40\% of the time and always smaller than the internal scatter when the target phases combinations $\bm{x_t}$ had the best fit model subtracted. Comparison stars (i.e. calibrating the calibrators from Table~1) also had small or zero values for $\Delta_t$ and calibrated closure-phase uncertainties of order 0.8 degrees for the $L'$ filter and 0.4 degrees for the $K'$ filter in each visit. These uncertainties are much lower than the uncalibrated closure-phases, which had typical median absolute values of $\sim$4 degrees for the L' filter and $\sim$2 degrees for the K filter. This highlights the need for a robust closure-phase calibration process.

It is difficult to directly show the quality of fits by directly plotting measured and model closure-phases. However, in the case of an image which is largely 1-dimensional, we can fit Fourier phase to closure-phase, where the phases are chosen by a least-squares minimisation process that both fit the closure-phase and attempts to fit the binary model (essentially filling in the missing phase information with the model). This is the approach of Figure 6 of Lacour et al (2011). The phases can then be plotted along baselines projected along the principle axis of the model image. We are able to do this for the K-band 2010 data, which has one dominant point source in our fitting, and show this in Figure~1.

Our closure phase image reconstructions used the Monte-Carlo MArkov Chain IMager algorithm (MACIM) \citep{Ireland:2006uq}. When using the mean image output, this algorithm is essentially identical to a maximum entropy method, and has been used many times in optical interferometry imaging  \citep{Monnier:2007fj,Zhao:2008kx}. The MACIM image model consisted of a point source with variable flux (i.e. free to be chosen by the algorithm) and an extended image. This point source represents the star, and without it as an explicit parameter in the model, maximum entropy like methods spread the central source flux throughout the image, especially when visibility amplitudes are poorly constrained. For our data, visibility errors are always at least a factor of $\sim$2 larger than closure-phase errors, which were typically less than 0.02\,radians. The visibility errors were also highly correlated and vary with adaptive optics Strehl ratio. Therefore, we chose to essentially fit only to the closure-phase data, and accomplished this by adding squared-visibility errors of 0.2 to the calibrated data. This in turn meant that we were insensitive to any point-symmetric extended flux in the image. We chose the field-of-view of the images to match the window function size in our data analysis, and chose the number of image elements so that the (super-)resolution of the final image was approximately $\lambda/2 D$.

In addition to producing reconstructed images, we also directly fit the closure phases with models described by a small number of point sources. This new multi-source fitting routine was motivated by high residuals seen in a fit with one companion (our standard technique), as well as by potentially complex structures seen in the reconstructed images. We first attempted to fit a single companion model directly to the closure phases by using a grid search, as has been standard for our previous observations. We then searched for solutions with up to 3 additional point source solutions in the vicinity of the best fit single-companion models. We used a gradient-descent least squares algorithm in up to 12 dimensions (3 contrasts each in $K'$ and/or $L'$, 3 separations, and 3 position angles) for this fitting, making use of the IDL mpfit library. As described above, we fit to statistically independent linear combinations of closure phases and have $\chi^2 \sim 1$ by construction (due to the $\Delta_t$ parameter), so we report the formal errors directly output by the mpfit program. The source(s) described here are located at $>$$\lambda /D$, so they are not subject to the degeneracy between contrast and separation that was seen for the similar detection of T Cha \citep{Huelamo:2011fk}.

We also attempted to reconstruct images based on the full complex visibilities (i.e., using both amplitude and phase) in order to retrieve information on the point-symmetric extended flux (or lack thereof). The 2009 $L'$ data and 2010 $K'$ data had amplitudes too noisy to be useful in this high-contrast regime, but the 2010 $L'$ data gave images consistent with the closure-phase images that we found. We also used point source fits (as described above) in order to more quantitatively measure whether most flux comes from a flux-symmetric component (such as a disk) or from the asymmetric component (a companion).

The calibrated oifits files \citep{Pauls:2005yq} and the code that produced the fits and images for this paper have been made available at \url{http://www.physics.mq.edu.au/$\sim$mireland/LkCa15\_sup/}.


 \begin{figure*}
 \epsscale{1.0}
 \plotone{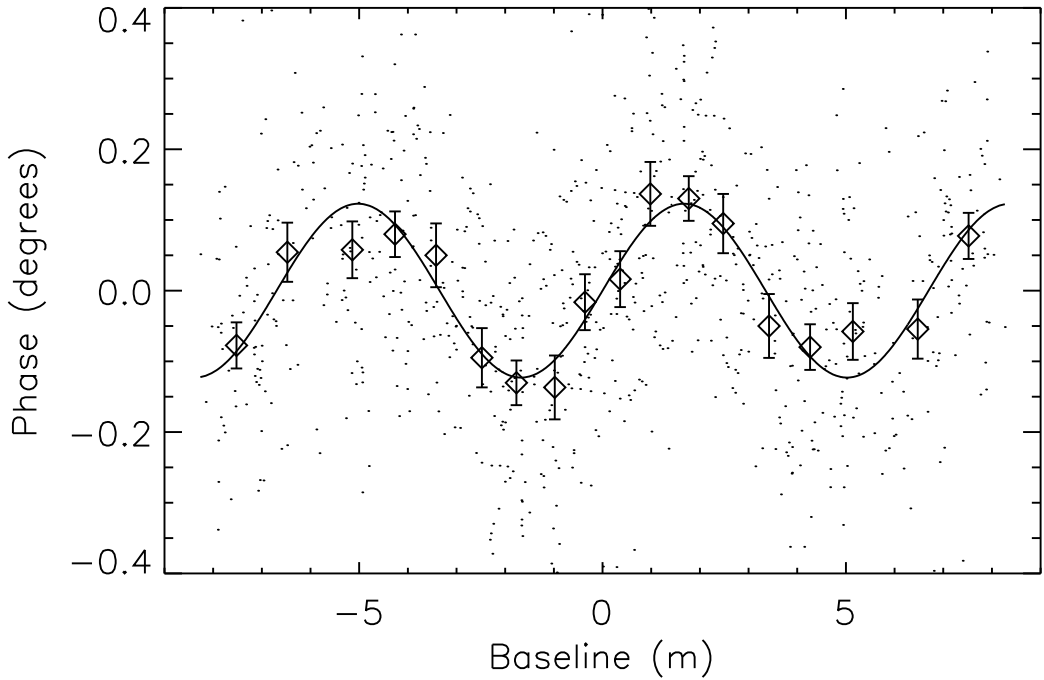}
 \caption{\footnotesize Fourier phase fitted to closure-phase (small dots) and the binned version of the same observable (triangles) for all 2010 K-band data on LkCa~15, plotted against the baseline projected along the principle axis of the best fit binary model. The phases of the best fit binary model model from Table~2 is shown as a solid line..}
 \end{figure*}


\section{Results}

\subsection{A (Proto)planetary Companion to LkCa 15?} 

Our $L'$ observations in November 2009 found a closure phase signal that was inconsistent with that of a single point source -- that is, a star with no companions -- at a confidence level of $>10\sigma$. We initially fit this source with a single faint companion, and found that it significantly improved the fit in both the discovery epoch and a followup $L'$ epoch in August 2010. However, the residuals in the fit remained consistently higher for both epochs than was seen for the calibrators, hinting that the ``companion'' could represent a more complex structure than a single point source. A third $L'$ observation in November 2010 confirmed that the system indeed required multiple point sources to fully explain its closure phases and visibilities. However, a deep observation in $K'$ further complicated the picture, as those visibilities suggested most of the flux was located at a single position in between two peaks of $L'$ flux.

In Figure 2, we show the results of independent image reconstructions of the closure phases from the 2009 November $L'$, 2010 August $L'$, 2010 November $L'$, and 2010 November $K'$ epochs. These data portray a system with complex color-dependent morphology. The $L'$ flux is seemingly dominated by two bright peak at similar projected separations and with position angles $\sim$50 degrees apart, while most of the $K'$ flux comes from a single point source located between the $L'$ sources. There is some evidence for emission in $L'$ between the two main peaks, though with less significance. The shorter $L'$ observations (from the 2010 epochs) show more blending of the $L'$ emission into a single elongated structure, which could indicate that these observations lack sufficient $S/N$ to support full image reconstructions. There is also a $K'$ peak near the position of the southwestern $L'$ peak, suggesting a possible counterpart for that component as well.

In Table 2, we list the corresponding astrometry and photometry derived from directly fitting the closure phases of each epoch (and various combinations of them) with models including 0, 1, 2, or 3 additional point sources. This approach offers a more quantitative and sensitive measure of the system's properties, and since the entire structure is only $\sim$3-4 times the size of the effective resolution, then a decomposition into point sources should encapsulate most of the useful morphological information. A mismatch between our model and the true complexity of the observed structure could result in systematic uncertainties, though, so the results should be treated with caution.

Each of the individual $L'$ epochs allows for two point sources at a statistically significant level ($\sigma < 0.2$ mag or $S/N > 5$), while combining at least two of the $L'$ epochs allows for the fitting of three statistically significant point sources. The $K'$ epoch also allows for a fit with three statistically significant point sources, albeit with the third source only barely significant. Given the overall morphology observed in the reconstructed images, we have classified each source as corresponding to a northeast (NE), central (CEN), or southwest (SW) component. Wavelength-dependent changes in the fit position (as for the projected separations of CEN in $L'$ and $K'$) suggest that the underlying morphology is indeed more complex, but further decomposition is not warranted by the resolution of our data. Even at this scale, it is possible that there is degeneracy between the fluxes of the flanking components and the central component, with flux from NE and SW contaminating the measurement for CEN. However, the positions of each component can be attributed to flux seen in the reconstructed images of Figure 2, suggesting that the overall morphology is being captured by both methods.

The $L'$ visibility amplitudes are too noisy to contribute to a full image reconstruction, and even point source fits are too noisy for the same level of detail as fits based on purely closure phases. However, in a fit to the 2010 $L'$ data with point-symmetric and antisymmetric models of three point sources (i.e., two companions), we find that the total flux is $1.7 \pm 0.3 \%$ using full visibilities and $1.4 \pm 0.1 \%$ using only the closure phases. This fit indicates that $91 \pm 9 \%$ of the total flux comes from the antisymmetric structures seen in the reconstructed images, with no more than the remaining 9\% coming from a point-symmetric component (such as a disk). This strong limit shows that our observed sources do indeed represent localized structures, rather than bright clumps embedded in a disk.

Given the unusual nature of this source, we must consider the validity of the detection. The use of inappropriate (i.e. binary) calibrators can lead to the detection of spurious sources. However, we can reject this hypothesis since the companion was detected on six different nights over two years, typically using different sets of calibrators. We also tested the observations by omitting each calibrator in turn to see if the detection remained. The detection lost significance due to the smaller number of calibrators remaining, but it remained at all epochs. This same degree of persistence also allows us to reject the possibility that we are seeing a background source (which given its spatially resolved nature, would need to be a multiple system or a galaxy). The proper motion of LkCa 15 reported by UCAC3 is (+9.6,-13.3) $\pm$ 2.2 mas/yr\citep{Zacharias:2010lr}, so if we were observing a background source or sources, then the $L'$ detections should have moved by ($\Delta \alpha$,$\Delta \delta$) = (+8.4,-11.6) $\pm$ 1.9 mas between the 2009 and 2010 epochs. The NE and CEN sources appear to yield low-quality fits at individual epochs, perhaps because flux is allowed to shift between the two nominal positions. However, the SW source appears fairly consistent between all epochs, and hence provides the best opportunity to measure the relative motion. We found that between 2009 (November) and 2010 (August plus November), the source position differed by ($\Delta \alpha$,$\Delta \delta$) = (-7.5,-8.8) $\pm$6.7 mas. The disagreement with nonmovement is therefore 16.2 $\pm$ 7.0 mas, or 2.3$\sigma$. Intriguingly, this motion is almost entirely in the PA direction, suggesting that we might be seeing orbital motion. We also note that the probability of such a chance alignment is extremely small; the 2MASS Point Source Catalog lists only 77 sources with $K<15$ (and hence $L\la$15) within a radius of $<$5 arcminutes of LkCa 15, for an overall density of $3 \times 10^{-4}$ arcsec${-2}$. The probability of finding one chance alignment within the disk gap of LkCa 15 ($\rho \la 300$ mas) is only $8 \times 10^{-5}$; even a survey of all $\sim$200 stellar members of Taurus would have only a very small probability of finding a chance alignment.

Given the observed morphology of the $L'$ and $K'$ observations, then the LkCa15 system appears to represent at least four sources: the primary, plus three sources of spatially resolved flux.  We illustrate this geometry in Figure 3, where we show a multicolor RGB image with superpositions of an $L'$ reconstruction based on all epochs (in red) and the $K'$ reconstruction (in blue), along with submm observations of the disk for context \citep{Andrews:2011lr}. With this geometry in mind, we have simultaneously fit the observed closure phases at all epochs to obtain the most reliable characterization of these components (Table 2, bottom section: ``Global Fit''). This fit indicates that there are detections of $K'$ emission at the sites of the $L'$ emission, and vice versa. However, since all of the sources are separated by the diffraction limit, then some caution is required; if the sources are more complex than simply 3 point sources, then a mismatch between the model and the observations could cause apparent flux to be transferred between the sources.


 \begin{figure*}
 \epsscale{1.0}
 \plotone{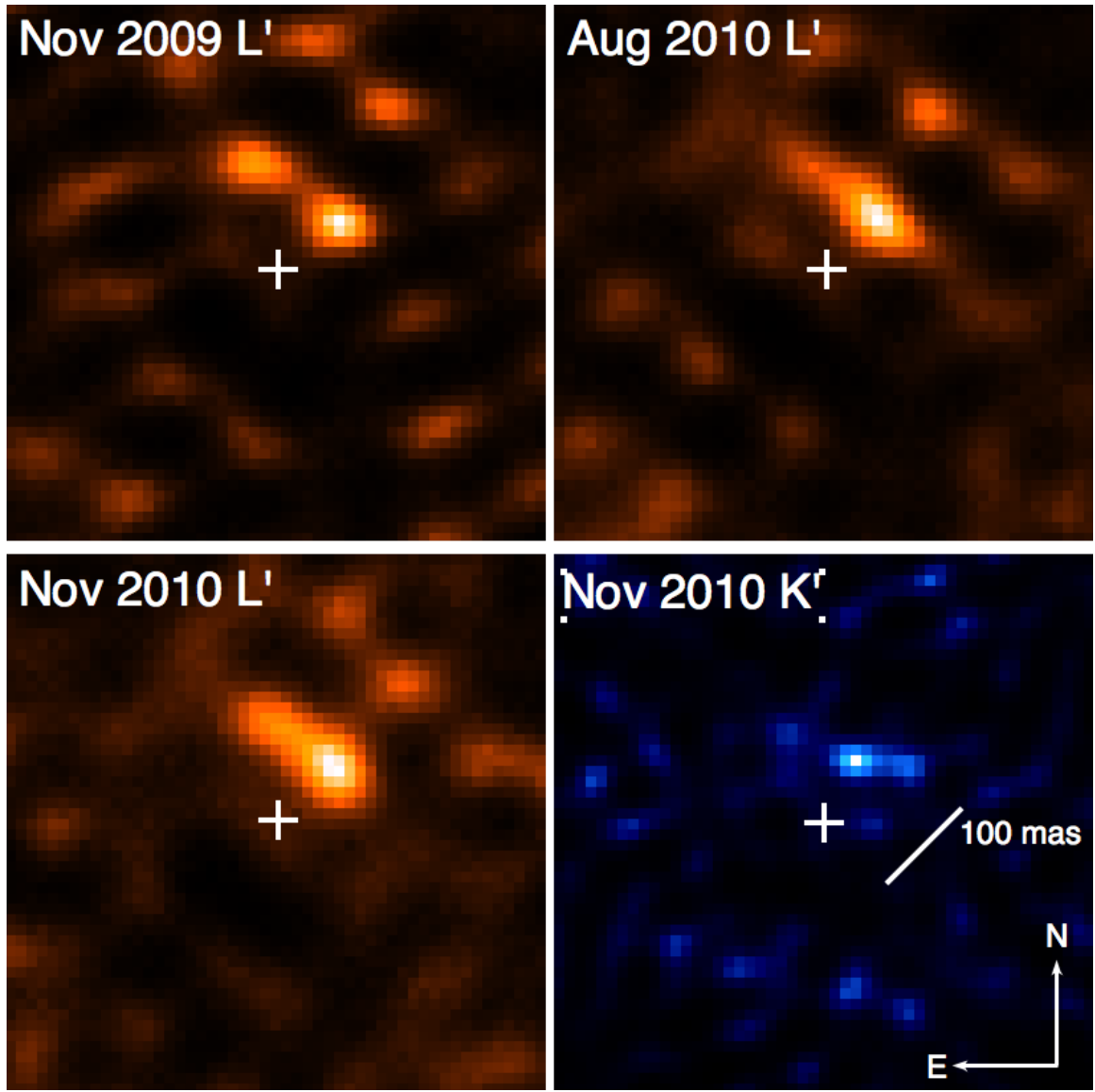}
 \caption{\footnotesize Reconstructed images for the four observations of LkCa 15: 2009 November $L'$ (upper left), 2010 August $L'$ (upper right), 2010 November $L'$ (lower left), 2010 November $K'$ (lower right). Each image was reconstructed with a pixel scale of 10 mas (total FOV = 0.5\arcsec), and the stretch was chosen to reveal the noise peaks (most of which are aliased power from the genuine detections) without saturating the statistically significant detections (see text). We denote the location of the central star with a cross and show a 100 mas bar for scale; all significant structures lie at or inside this angular distance from the central star. The shorter observations in $L'$ (from 2010 August and 2010 November) show more smearing of the spatially resolved structure, indicating that the data quality is not quite sufficient for image reconstruction to be effective. However, direct fits to the closure phases (Table 2) show that the fits with a set number of point sources are typically consistent to within the uncertainties.}
 \end{figure*}


 \begin{figure*}
 \epsscale{1.0}
 \plotone{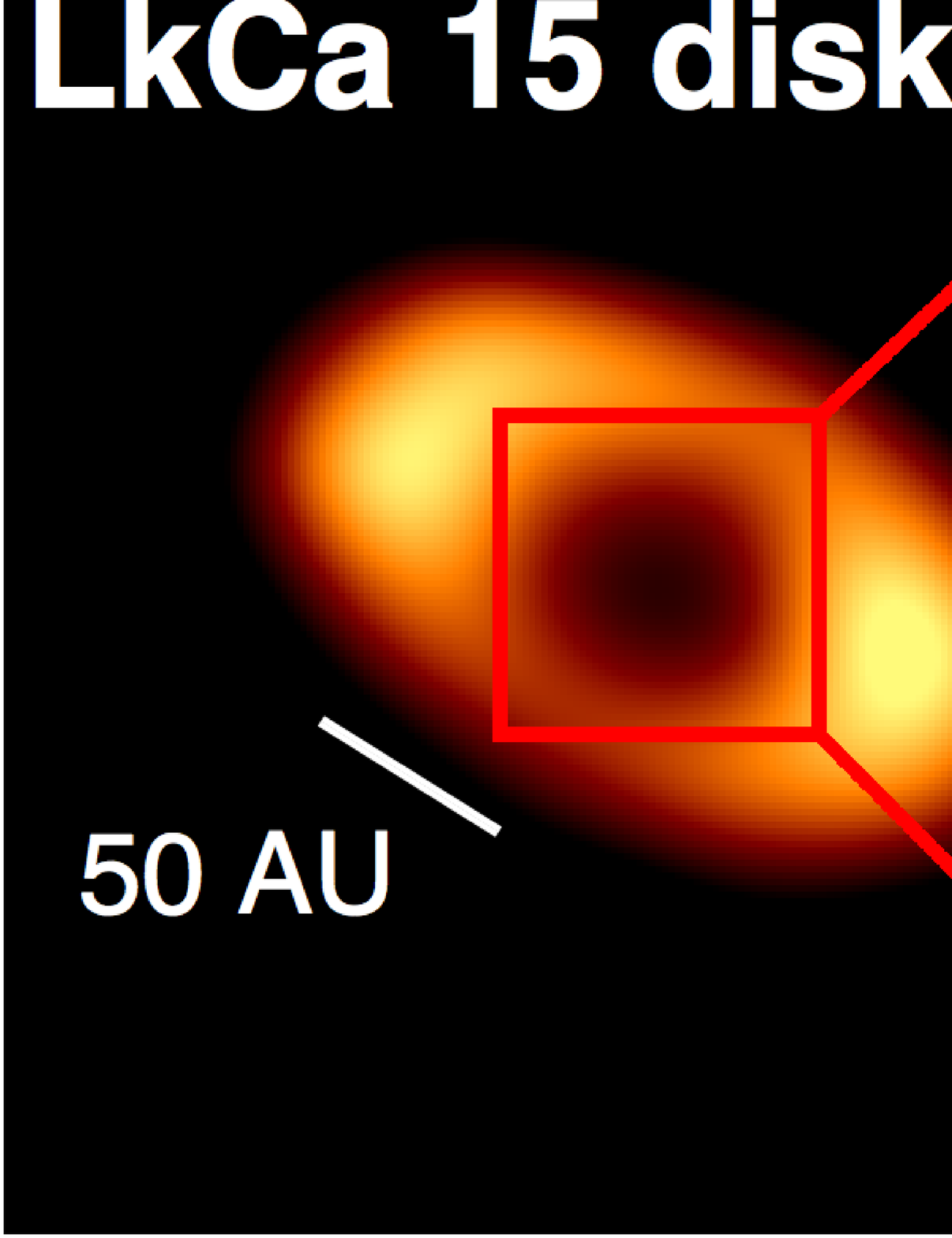}
 \caption{Left: The transitional disk around LkCa15, as seen at a wavelength of 850 $\mu$m (Andrews et al. 2011). All of the flux at this wavelength is emitted by cold dust in the disk; the deficit in the center denotes an inner gap with radius of $\sim$55 AU. Right: An expanded view of the central part of the cleared region, showing a composite of two reconstructed images (blue: $K'$ or $\lambda = 2.1$ $\mu$m, from November 2010; red: $L'$ or $\lambda = 3.7$ $\mu$m, from all epochs) for LkCa 15. The location of the central star is also marked. Most of the $L'$ flux appears to come from two peaks that flank a central $K'$ peak, so we model the system as a central star and three faint point sources.}
 \end{figure*}


\subsection{Orbital, Morphological, and Atmospheric Properties} 

The observed morphology of LkCa15's candidate companion is more complicated than that of older directly-imaged exoplanets, which are seen as unresolved point sources \citep{Marois:2008zt,Kalas:2008jl,Lagrange:2009fc}. The flux is mostly concentrated in a single unresolved location at 2.1$\mu$m, but it is clearly extended at 3.7 $\mu$m. The most simple interpretation is that the central source is therefore a newly-formed exoplanet, which emits significant flux at 2.1$\mu$m due to either a warm atmospheric temperature or accretion of hot material. The surrounding 3.7$\mu$m dominated emission would then trace extended circumplanetary material, most likely as it is accreting down to the planet, though perhaps as it accretes past the planet and onto the inner disk \citep[e.g.,][]{Dodson-Robinson:2011fk}. We can extrapolate the orbital radii, absolute magnitudes and colors of these structures from our global fit of all observations (Table 2, bottom section) using the apparent magnitudes, distance, and age for LkCa 15 that we describe further in Appendix A.

We converted the observed separation and PA for each source into a deprojected orbital radius using the observed disk geometry ($i=49^o$, PA$=241^o$) \citep{Andrews:2011lr}: $R_{NE} = 20.1 \pm 2.8$ AU, $R_{CEN} = 15.9 \pm 2.1$ AU, $R_{SW} = 18.4 \pm 2.6$ AU. Model fits for disks typically vary by $\sim$5--10$^o$ between different observations and models of the same targets, so we adopt a systematic uncertainty of $\pm$5$^o$ in the inclination; combined with the distance uncertainty of $\pm$15 pc, the total uncertainty in deprojected radii is $\sim$15\%. Given deprojected orbital radii of $\sim$16--20 AU, then the corresponding orbital period and orbital motion around a solar-type star are $\sim$90 years and $\sim$4 deg/year. Our astrometric precision for the central source (i.e., the proposed planet itself) is $\sim$1.5$^o$ (for its $K'$ emission), so it is plausible that we could see orbital motion at the 3$\sigma$ level within the next 1-2 years. Orbit determinations for other high-contrast companions (such as GJ 802 B; Ireland et al. 2008) show that the astrometric errors predicted by NRM are typically valid. The $L'$ astrometry for the SW source might already be showing orbital motion, since the offset between 2009 and 2010 is almost entirely in the PA direction and has a magnitude of 1.7$\sigma$. However, if the emission comes from a spatially resolved region, then it could be subject to two uncertainties. Since we are fitting a potentially resolved source as a point source, model mismatch could cause systematic astrometric errors. More seriously, if the emission comes from an extended dusty structure, then the centroid of the emission itself could change (with respect to that structure's position) over time. Even if the dust producing the $L'$ emission is orbiting at a Keplerian velocity, the emission from different points in the structure might wax or wane. A conservative estimate of orbital motion should be based on at least several additional epochs, in order to determine the residuals around its apparent orbital velocity.

The observed contrasts can be converted into absolute magnitudes using the observed photometry for LkCa15 A (Appendix A) and the distance to Taurus-Auriga, 145$\pm$15 pc \citep{Torres:2009ct}; the combined absolute magnitude and color for all three components are $M_{L'}=6.8\pm0.2$ mag and $K'-L'=1.7 \pm 0.2$ mag. Young hot-start planets should have SEDs similar to L dwarfs, so assuming an approximate temperature of 1500 K and appropriate bolometric corrections \citep{Leggett:2002qz}, then the corresponding bolometric luminosity is $L_{bol} = 2 \times 10^{-3}$ $L_{\odot}$, with an uncertainty of at least a factor of 2--3 (depending on the actual temperature). 

Since the observed flux comes from spatially resolved structures and not a single point source, then the physical properties of each component must be considered individually.  If the flux seen from the central source (near the $K'$ peak) corresponds to the planet, then its brightness and color ($M_{K'}=9.1 \pm 0.2$; $K'-L'=0.98 \pm 0.22$) are more consistent with a photosphere than with warm dust. For ages of 1 Myr or 5 Myr (bracketing the 1$\sigma$ limits on the age of LkCa 15), then this brightness would naively be consistent with a mass of 6 $M_{Jup}$ or 15 $M_{Jup}$ according to the ``hot start'' models \citep{Chabrier:2000sh}. However, if this planet is newly formed, then even the value for 1 Myr might be an overestimate. Furthermore, the presence of significant circumplanetary material suggests that it is quite likely to be accreting, and current planet formation models suggest that a giant planet should intercept much of the disk mass that would otherwise accrete onto the central star \citep{Lubow:2006fk,Machida:2010lr}, typically $\dot{M}=$$10^{-7}$--$10^{-9}$ $M_{\odot}/yr$ \citep{Gullbring:1998pt}. For an accretion rate of $\dot{M}=$$10^{-8}$ $M_{\odot}/yr$, the corresponding accretion luminosity would be $\sim$10$^{-3}$ $L_{\odot}$, assuming an emission temperature of 1500K (consistent with the observed $K-L'$ color), a planetary mass of 5 $M_{Jup}$, and a planetary radius of $<$5 $R_{Jup}$. Accretion therefore could explain all of the observed luminosity, and since a reservoir of 55 $M_{Jup}$ of material remains in the outer disk, then the proposed accretion rate of $10^{-8}$ $M_{\odot}/yr$ could be sustained for the entire 5 Myr lifetime of a typical protoplanetary disk.

The emission from the surrounding material appears to be much redder (and hence cooler), with much more flux emitted at 3.7$\mu$m than at 2.1$\mu$m ($M_{L'}=7.5 \pm 0.2$ mag and $K-L' = 2.7 \pm 0.3$ mag for NE, $M_{L]}=7.4 \pm 0.2$ mag and $K-L = 1.94 \pm 0.16$ mag for SW). This red color suggests that the material is quite cool ($T_{eff}<$1000 K), though still much warmer than the ambient temperature at this distance from the star ($\sim$100 K for large dust grains; Section 3.4).  The presence of circumplanetary material is expected, since both the planet and inner disk should be accreting mass from the remaining protoplanetary disk (where a substantial reservoir of material remains, 55 $M_{Jup}$). However, the material is much more spatially extended than models have predicted \citep{Machida:2010lr}. Our observations place the emitting material at 6$\pm$1 AU away from the central source, which is larger than the Hill radius for a 10 $M_{Jup}$ object (2--3 AU) and significantly larger than the radius at which material heated by the planet should be emitting in the NIR ($<<$1 AU). Conversely, the distance is too small for the flanking components to represent material accumulating at the Trojan points. These two components are separated from the central component by only $20^o \pm 3^o$ in the deprojected plane of the disk, whereas the Trojan points should lead and trail a planet by 60$^o$.

Given the luminosity of LkCa 15, micron-sized dust grains at an orbital radius of $\sim$25 AU should have an equilibrium temperature of only 100 K, which is too cool to produce significant flux at 3.7$\mu$m. This suggests that energy is being generated (or delivered) to the extended circumplanetary environment in some other way. The direct accretion luminosity of the planet is 3 orders of magnitude smaller than the luminosity of LkCa15 A, so direct irradiation by the planet is also insufficient. Another plausible explanation is the deposition of orbital kinetic energy as material accretes into the circumplanetary environment from the outer disk. The expected accretion rate and typical orbital velocities at that radius from the star are large enough to deliver the needed energy, but detailed modeling will be needed to determine if material will be heated sufficiently as to emit in the NIR. Finally, one possible explanation is that the energy is transported out from the planet by accretion jets or winds, as a significant fraction of accreting material should be launched back outward from the planet \citep{Herbig:1950zr,Haro:1952mz,Konigl:2000ly,Shu:2000gf}. If this higher-velocity material impacts the complex circumplanetary environment, perhaps guided by the global magnetic field of the disk, then it could deposit sufficient energy to heat that environment. More rigorous testing of these models will required additional observations at shorter or longer wavelengths (in order to refine the temperature estimates for each spatially resolved component) or direct identification of the circumplanetary dust distribution with observations from ALMA.

In Figure 4, we show the brightness and color of the individual components and the full structure as compared to free-floating stars and brown dwarfs within the Taurus star-forming region \citep{Luhman:2010cr}. The observed fluxes for any individual component, or even for the sum, are fainter than all but the few least-massive members of Taurus-Auriga, which themselves fall in the planetary-mass range \citep[e.g.,][]{Luhman:2009wd}. If other explanations can be rejected, then this low luminosity strongly suggests that our observations have revealed a planetary companion.


 \begin{figure*}
 \epsscale{0.85}
 \plotone{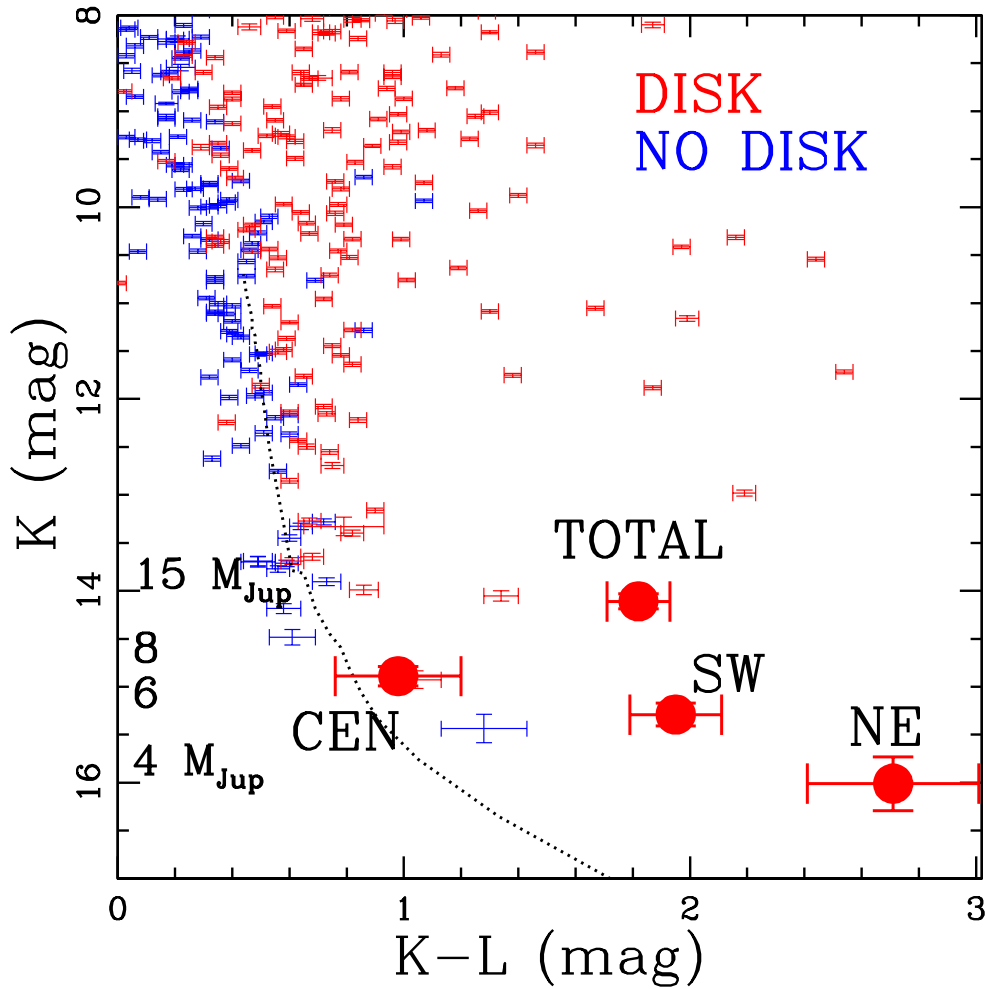}
 \caption{(K, K-L) color-magnitude diagram for Taurus, showing all free-floating members of Taurus \citep[open error bars;][]{Luhman:2010cr}, our measurements for the three distinct morphological segments of the resolved structures (labeled NE, CEN, and SW), and the combined color and magnitude of all three (TOTAL). Taurus members without disks are shown in blue, and tend to follow the 1 Myr theoretical isochrone of the Lyon models (dotted line) \citep{Chabrier:2000sh}. Objects with circumstellar disks tend to be redder than those without. The central source also falls near the 1 Myr isochrone, suggesting that it could represent the actual planet; if so, the inferred mass is $\sim 6 \pm 1$ $M_{Jup}$. However, the luminosity for this source could be dominated by accretion luminosity and hence not indicative of the true planetary mass.}
 \end{figure*}


\subsection{Heating Processes for The Extended Circumplanetary Material} 

We interpret our observations to represent a single planet (with a relatively neutral near-IR color) surrounded by resolved circumplanetary material (with a very red near-IR color). The modest color of the planet is not surprising, since its brightness is likely dominated by accretion luminosity that should be much bluer than the underlying photosphere. However, the brightness and size of the circumplanetary region requires additional explanation. As we discuss further in Section 3.4, material should only glow in the NIR if it has a temperature of $\ga$500--1000 K; otherwise, its blackbody peak shifts to much longer wavelengths. Even for relatively luminous stars, dust is only heated to this temperature within the inner $\sim$0.1--0.2 AU \citep{Olofsson:2011lr}. We therefore must question if radiative heating from a planet is sufficient. Given the $K'$ magnitude and $K'-L'$ color of the proposed planet, then its total luminosity is $L \sim 10^{-3} L_{\odot}$. Large dust grains should absorb this incident flux in proportion to their cross-section and emit it in proportion to their surface area, so their equilibrium temperature depends only on the distance from the planet. At a distance of $\sim$5 AU, the equilibrium temperature for large dust grains should be $\sim$20--25 K, a factor of 20 too low to explain the observed $L'$ flux.

This energy could also represent shock-heating of circumplanetary material, due to the deposition of orbital kinetic energy from disk material into the circumplanetary environment. The typical orbital velocity at an orbital radius of $\sim$20 AU from LkCa15 A should be $\sim$7 km/s, and the kinetic energy of accreted disk material could be liberated as thermal energy if it impacts on circumplanetary material moving with a similar speed, but in a different direction. If the typical accretion rate is $\sim$10$^{-7}$--10$^{-8}$ $M_{\odot}/yr$, then the corresponding luminosity will be $\sim$10$^{-3}$--10$^{-4}$ $L_{\odot}$. The high end of this range is consistent with the observed luminosity, so this explanation could be feasible. However, the observed temperature of such material will depend on the detailed physics of the shock-heating process, and hence will require more sophisticated modeling and analysis. Furthermore, the cooling timescale for small dust grains is very short ($<$1 sec), so the similarity between our 2009 and 2010 observations suggests that this heating is a continuous process and not the result of a single energetic event (such as a collisional cascade following a collision of large planetesimals).

Another plausible explanation is that we are indeed seeing energy from the final accretion of material onto a central planet, but it is being transported to the observed sites ($\sim$5 AU away from the planet) through some process other than direct radiation that is not attenuated by a factor of $r^{-2}$. Accretion onto young stars is inextricably tied to the launching of jets \citep{Herbig:1950zr,Haro:1952mz} and winds \citep{Konigl:2000ly,Shu:2000gf}, and those winds and jets can carry a significant amount of the accreted material outward at a high velocity. If planetary accretion also feeds such outflows, then this high-speed material could impact cool material in the surrounding circumplanetary environment and transform its kinetic energy into thermal energy, forming planetary equivalents of the bow shocks observed around stellar outflows. However, this explanation invokes even more open topics of star and planet formation than the previous hypothesis, so it also must await testing with more sophisticated modeling.

Finally, a consistent (but improbable) explanation for the geometry might be that we are seeing two extremely red planets, and that the central blue feature represents a non-planetary feature. Given the extreme youth of the system, it is plausible that a stable orbital solution exists for two similar-luminosity (and hence similar-mass) planets to share nearly the same orbital radius for a sufficient length of time. If these planets were locked into a 1:1 resonance, then they might also support a quasistable point (equivalent to Trojan points) midway between them. If dust accumulated at this quasistable point, then it would reflect light from the central star with approximately the stellar color \citep{Duchene:2004qy}, matching the observed color of the central source. The most expedient way to rule out this improbable solution is to widen the wavelength coverage of our observations (i.e., to $H$ or $M'$), which should distinguish intrinsically cool planets from intrinsically blue reflected light.

\subsection{Rejected Alternate Explanations} 

We also must weigh alternate explanations for our observations. As we discussed in Section 3.1, the consistent astrometry across a one-year baseline suggests that our discovery is astrophysical and is comoving. One possibility is that the structure could also be caused by a more massive companion that is obscured at the wavelength of our observations, and hence made to look fainter. Since LkCa 15 is surrounded by a dusty disk, then we could also have observed thermal emission from dust that is heated by the primary or directly reflected light from the primary that is incident on the disk. Finally, we could be observing line emission from the gas (i.e., Br$\gamma$) or from PAHs.

The purported planetary companion could represent an unseen binary companion, perhaps obscured by an edge-on disk so that it appears significantly fainter. The example of CoKu Tau/4 demonstrates that ``transitional disks'' can host binary companions \citep{Ireland:2008kx}, and other binary systems, like HK Tau and HV Tau, are known to have wider components which are obscured in this way \citep{Stapelfeldt:1998iv,Stapelfeldt:2003nt}. However, most close binary systems are thought to have coplanar disks \citep{Jensen:2004ux}, so any disk around this putative stellar companion should not produce significant obscuration along our line of sight. The observed geometry also is inconsistent with that seen for edge-on disk systems like HV Tau \citep{Duchene:2010xt}, where the central dust lane also separates the flux into separate lobes,  but those lobes are more widely separated at shorter wavelengths (due to the higher extinction through the disk near the midplane). Finally, if the putative companion was obscured by a massive circumstellar disk, then it should produce significant mid-infrared and submm continuum emission that would be distinguished from the outer disk \citep{Espaillat:2007rq,Andrews:2011lr}.

We also must consider the prospect that the flux we attribute to a planet could instead represent energy from the central star that is being processed and re-emitted by the disk. The central hole is thought to be largely cleared of disk material, but it is plausible that grain growth and settling could instead cause the low apparent optical depth for submm photons. However, at orbital radii of $\sim$20 AU from a star with $T_*=4000$\,K and a stellar radius of $\sim$2\,$R_{\odot}$, the equilibrium temperature is only $\sim$100 K, such that the blackbody peak is at $\sim$30 $\mu$m; emission in the $L'$ filter would be negligible in comparison. This simple calculation is consistent with more complex models of disk structure and evolution, which find that only the innermost regions reach temperatures of $>$500 K \citep{DAlessio:2006qy}. In fact, for a full optically thick disk, the majority of light that is processed and re-emitted in the near-infrared comes from the inner $\sim$0.1 AU of the disk \citep{Olofsson:2011lr}. Even inner gaps with a size of $\sim$1 AU lead to an observed paucity of flux at $<$10 $\mu$m \citep{Najita:2010vn}. Small dust grains can be much warmer than large dust grains, but even they can not achieve the necessary temperature. Where gray dust has an opacity proportional to $\lambda ^{-1}$, the dust temperature is given by:

\begin{equation}
T_d = (r_d/2R_*)^{2/5} T_*
\end{equation}

And hence the dust temperature could be as high as 246\,K. With an optical depth of unity at the Planck peak of 700\,nm and a 25 AU$^2$ cross-sectional area (the maximum allowed by the resolution of Keck), even this extreme dust could not provide sufficient flux to match the point sources we model in L-band. Furthermore, this hypothetical dust would shadow the outer disk, which is not observed, and it can not explain the $K'$ morphology.

Another possibility is that we are seeing light from the disk, but rather than thermal emission from heated material, it is direct reflection of incident light from the central star. Reflected light has been cited as the likely explanation for the ring of circumstellar light previously reported in $H$ band \citep{Thalmann:2010mz}, which is coincident with the $\sim$50 AU radius of the inner disk edge suggested by SED modeling \citep{Espaillat:2007rq,Espaillat:2010xj} and directly observed in submm emission \citep{Andrews:2011lr}. The disk geometry suggested by the $H$ band emission somewhat favors placing the northwest face of the disk in the background, with a nearly normal incidence of reflection off the face of the disk wall at $\sim$50 AU. If there were another wall located at a radius of $\sim$20 AU, then it could reflect light in a similar manner. However, this explanation faces a significant challenge. Any inner wall at smaller separations must reflect $\sim$1.5\% of the incident $L'$ stellar flux from an arc of $\sim$60$^o$, or $\sim$10\% of the incident flux in a complete circle around the star. The wall height needed to intercept this amount of flux ($\sim$5 AU) would shadow the wall at 50 AU, an effect that is not seen in the SED fits or the reflected light. This explanation might still be feasible if a wall at 25 AU were optically thick at $L'$, but optically thin at optical wavelengths (allowing the majority of the stellar flux to pass). However, this type of opacity law is very much contrary to all standard dust types, which are either grey or have opacity increasing to shorter wavelengths \citep{Schlegel:1998yj}.

If the disk geometry were reversed, placing the northwest face in the foreground, then our reported detection could also represent forward-scattered light from the central star, which is being reflected back to our line of sight by material inside the disk. As for the previous case, we must give this explanation extra merit since we have detected the spatially resolved flux near the minor axis of the disk, as we would expect from forward scattering. However, the same drawbacks also apply, in that any material must reflect sufficient light at $K'$ and $L'$ to explain our detection, but still remain optically thin in the optical. Furthermore, the extremely red color of the light is also difficult to explain with forward scattering. Most observations of disks in scattered light, such as for the circumbinary ring of GG Tau \citep{Duchene:2004qy}, find colors which are neutral or moderately blue. The extremely red color of our detection is not consistent with these other observations. Even if the dust grain size distribution were chosen to optimally redden the $K'-L'$ color (i.e. a single-sized grain population with a radius of $\sim$1.5 $\mu$m), then it could only redden the reflected light by $\Delta (K'-L') \sim$0.5 (to $K'-L'\sim$1.5), whereas the red sources have an observed color of $K'-L'=$2.0--2.7.

Finally, we must consider the possibility that we are not observing continuum emission from a planetary companion and circumplanetary dust, but instead are seeing line emission from gas (via Br$\gamma$ at 2.16 $\mu$m) and PAHs (at 3.3 $\mu$m). However, the blue edge of the $L'$ filter is located at 3.43 $\mu$m, so only the extreme red wing of the PAH line might pass any light through this broadband filter. Furthermore, if all of the flux observed were line emission, then given the apparent broadband flux ratios ($\Delta K' = 6.7$ and $\Delta L' =  4.7$) and the ratio of the line width to the width of the full broadband filter (e.g., $>>$100:1 for Br$\gamma$), the line emission should exceed the continuum and be observed clearly in spectra of the entire system. Previous observations at 2--5 $\mu$m have seen no evidence of such line emission \citep{Espaillat:2008eq}. Nonetheless, such emission would produce a large spectroastrometric signal \citep[e.g.,][]{Pontoppidan:2011lr}, so such an observation should be attempted to conclusively rule out this explanation.

\section{Implications for Planet Formation and Evolution} 

The apparent planetary companion to LkCa 15 is the first likely exoplanet to be discovered at its time of formation, and hence it provides a new view of planet formation and early planet evolution. Planet formation models make unique predictions regarding the location and epoch of planet formation \citep{Pollack:1996dk,Boss:2001sd,Ida:2004vh}, so the orbital radius of the planet and the age of its parent star provide the first direct evidence for distinguishing between these models. Also, evolutionary models make extremely discrepant predictions regarding the luminosity of planets as a function of age \citep{Chabrier:2000sh,Fortney:2008lo}, though this discussion is complicated by possible flux contributions from circumplanetary material and accretion luminosity.

If the planet is coplanar with its disk, then the current orbital radius is 15.7$\pm$2.1 AU. The core accretion model of planet formation is primarily limited by the ability to assemble a 20 $M_{\oplus}$ core, so it is thought to form planets much more efficiently near the snow line ($a\sim$3-5 AU) than at these larger radii \citep{Tsiganis:2005fk,Kennedy:2008fk}. Recent models of in-situ formation suggest that a Saturn analog could potentially form at $a\sim$10 AU within $\sim$3--4 Myr \citep{Dodson-Robinson:2008pb}, but for Uranus and Neptune analogs to be formed within $<$5 Myr, they must begin at $a<$15 AU and then subsequently migrate outward \citep{Dodson-Robinson:2010qy,Bromley:2011lr}. In contrast, disk instability models are more efficient at forming planets at larger radii ($a>$25--50 AU) \citep{Rafikov:2005ys,Boley:2009or,Meru:2010rt,Boss:2011fk} since they are primarily limited by the ability of gas to efficiently cool and by the shear of differential rotation speeds, both of which limit the ability for a protoplanetary clump to exceed the Toomre stability criterion \citep{Toomre:1964fr}. The large orbital radius is therefore more consistent with formation via disk instability, though it is unclear why fragmentation occurred at a radius of 20 AU, rather than in the massive, cold outer disk. It is possible that orbital migration has already occurred, and hence planet formation occurred at either larger or smaller radii.

The implications of the age of LkCa15 are not as clear. A comparison to stellar evolutionary models suggests that the most likely age for LkCa15 A is 2 Myr, with a 1$\sigma$ range of 1--4 Myr \citep{Kraus:2009fk}. As we discussed above, even formation at $a<$15 AU seems to require at least 2--3 Myr for core accretion models, so they would more plausible if the primary star were to fall at the older end of this allowed range. Formation via disk instability does not necessarily carry a strong age constraint, as the collapse interval is thought to be quite short. However, disk instability should be most likely at young ages, when the disk is most massive. It is unclear why a planet would only form after several Myr, when part of the disk mass has already accreted to the star, so if the planet is forming via disk instability then the star should fall at the younger end of the allowed age range. The distinction between these possibilities would grow stronger if the age of LkCa15 could be determined more precisely, such as by refining its temperature (and hence its position on the HR diagram). However, systematic uncertainties in the ages of young stars probably limit any such determinations until the models can be better calibrated \citep{Hillenbrand:2004bh}.

The distance between the planet and the inner disk edge is also somewhat at odds with theoretical expectations. When a planet clears a gap in its disk, then the size of the gap should be approximately equal to the planet's Hill radius \citep{Crida:2006zr}; even if the planet is $\sim$10 $M_{Jup}$, then its Hill radius is only $\sim$2--3 AU. However, models of the disk SED and direct submm observations of the dust distribution show that the inner edge of the disk has a radius of $\sim$55 AU \citep{Espaillat:2007rq,Andrews:2011lr}, which is $>$25 AU outside of the planet's current orbital radius. This discrepancy could indicate that there are other, less luminous planets in the system and we simply can not detect them yet, a possibility that has been suggested to explain the wide cleared gaps for many transitional disks \citep[e.g.,][]{Zhu:2011lr,Dodson-Robinson:2011fk}. However, it is also possible that the planet is on an eccentric orbit, and hence the radius of the inner disk edge is set by the apastron distance. This assertion will be difficult to test since the planet's likely orbital period is so long ($\sim$90 yr for a circular orbit), making a full orbit fit difficult for at least several decades. However, scattered-light observations of the disk's inner edge have shown that it is off-center compared to the primary star, which is often a sign of perturbation by a body on an eccentric orbit \citep{Thalmann:2010mz}. Additional modeling of the planet-disk interaction could provide constraints on the most likely orbit for the planet.

Finally, the luminosity also provides a crucial first datapoint for calibrating the brightness evolution of young giant planets. Models that include realistic simulations of the core accretion process (the ``cold start'' models) \citep{Marley:2007qx,Fortney:2008lo} suggest that even young exoplanets should not not more luminous than $\sim 10^{-5}$ $L_{\odot}$. By contrast, models which assume a higher-entropy initial state as for collapse out of a gas cloud (the ``hot start'' models) \citep{Chabrier:2000sh} predict that young planets could be as bright as $\sim 10^{-3}$ $L_{\odot}$. The spatially extended morphology that we observe makes it difficult to infer a strong measurement of the total luminosity for the planet,  but as we show in Section 3.2, the observed flux ratio corresponds to a luminosity of $\sim$10$^{-3}$ $L_{\odot}$. This luminosity falls at the top end of the range permitted by the ``hot start'' models, and several orders of magnitude higher than the estimates of the ``cold start'' models. The same result has been found for moderately older direct-imaged exoplanets, such as Beta Pic b \citep{Lagrange:2009fc,Quanz:2010mb,Currie:2011fk} and HR 8799 bcde \citep{Marois:2008zt,Bowler:2010vn,Marois:2010ee,Currie:2011pi,Madhusudhan:2011fk}.

However, there is a strong caveat regarding the luminosity we infer for LkCa 15's planetary companion. Since our survey is targeting planets in transition disks, then by definition, we have observed this apparent planet at a point where its luminosity should be maximized. A gap in the disk should only form once a gas giant planet has a significant envelope, which only happens when it is accreting a significant amount of the mass crossing its orbit into the inner disk. Observations of other young solar-type stars shows that material in their disks will typically move inward at an accretion rate of $\sim$10$^{-8}$ $M_{\odot}/yr$ \citep{Gullbring:1998pt}; the planet is therefore likely to be accreting material very quickly, and hence generating a significant accretion luminosity. Realistic values for the planet mass/radius and accretion rate yield accretion luminosities as high as $\sim10^{-3}$ $L_{\odot}$ (Section 3.2), matching our inferred value. This brief period of high luminosity is seen as a luminosity spike in the ``cold start'' models, which find that even a $\sim$1 $M_{Jup}$ planet could reach a peak luminosity of $\sim 5 \times 10^{-3} L_{\odot}$ at the epoch of peak growth. It therefore seems plausible that we are not seeing any flux from the planetary atmosphere itself, but merely the accretion excess that is being released by its rapid assembly.

The blue color of the central source and the red color of the surrounding material further supports this view. If the planet is rapidly accreting mass from the broader circumstellar environment, then it should be enshrouded in material that is in the final stages of accretion. For a free-floating brown dwarf, this material organizes itself into a disk \citep{Scholz:2007it}. However, those free-floating objects only accrete at a rate of $\sim$10$^{-13}$ $M_{\odot}/yr$ \citep{Herczeg:2009kx}, which is five orders of magnitude lower. It is unclear whether the same geometry could be sustained with so much mass being continuously added to the environment. It seems equally probable that the accretion would take place via a very inflated disk or through an unexpected geometry. If so, then the very red color of the surrounding material should only be expected. The central planet would appear relatively blue due to accretion, but our expectations for the surrounding environment are largely unconstrained by theory.

\section{Summary}

We have reported the direct-imaging discovery of a likely (proto)planet around the young transitional disk host LkCa 15, located at the middle of the known gap in its disk. Our observations have revealed a faint and relatively blue point source, surrounded by co-orbital emission that is red and resolved into at least two sources. The most likely geometry consists of a newly-formed gas giant planet that is surrounded by dusty material, and which has been caught at its epoch of formation. This discovery is the first direct evidence that at least some transitional disks do indeed host newly-formed (or forming) exoplanetary systems, and the observed properties provide crucial insight into the gas giant formation process.

Additional studies of this system, both theoretical and observational, will be necessary to more fully understand its complicated colors  and morphology, and ultimately to confirm its planetary nature. Broadband photometry at additional wavelengths should be feasible with existing instruments and will extend our knowledge of the spatially resolved broadband SED. Next-generation instruments like GPI will also be capable of observing LkCa15 with NRM, and they will yield low-resolution spectra that show any influence of broad molecular absorption bands due to water or methane. Finally, submm/mm observations with ALMA will directly track the dust in the LkCa15 system with sufficient sensitivity and resolution to distinguish the circumplanetary material, resolving any remaining ambiguities as to its spatial distribution and mass.
 
 \section{Appendix: The Properties of LkCa 15 A}

The young solar analog LkCa 15 is a K7 star located in the nearby (145 $\pm$ 15 pc) Taurus-Auriga star-forming region \citep{Kenyon:1995dg}. Based on its position in the HR diagram, it has an age of $2^{+2}_{-1}$ Myr \citep{Kraus:2009fk}. The primary star's mass, 0.97$\pm$0.03 $M_{\odot}$, has been measured from the rotation curve of the circumstellar disk \citep{Simon:2000ty}. It has an observed brightness of $m_{[3.6]}=7.61 \pm 0.05$ mag in the IRAC 3.6 $\mu$m filter, and thus is of similar brightness in the $L'$ band \citep{Rebull:2010xf}. According to 2MASS, its brightness is $K = 8.16 \pm 0.02$ mag \citep{Skrutskie:2006nb}. Given the well-known variability of young stars, the real uncertainty is probably at least $\sim$0.1 mag in each filter\citep{Carpenter:2002yu}. The extinction to LkCa 15 is negligible for the purposes of NIR observations, $A_V<1$ or $A_K<0.1$ \citep{Kenyon:1995dg}.

We previously observed LkCa 15 with NRM in the $K'$ band ($\lambda=2.1$ $\mu$m) to determine if this cleared region could indicate the presence of a binary companion, but found no companions with contrast $\Delta K \le$6.2 mag ($M > 12 M_{Jup}$) at a confidence level of 99.9\% or $3.3 \sigma$ \citep{Kraus:2008zr}. We have reanalyzed our $K'$ band data for LkCa 15 in order to place a more strict upper limit on the brightness of a companion at the known position from our $L'$ band detections. Our new analysis suggests a stronger limit of $\Delta K < 6.6$ mag, suggesting that our old dataset nearly detected the $K'$ band counterpart that we observed in November 2010.

\acknowledgements

We thank Sean Andrews, Gregory Herczeg, Michael Liu, Jonathan Williams, Lucas Cieza, Lynne Hillenbrand, and Peter Tuthill for helpful discussions regarding this manuscript. We especially thank Sylvestre Lacour for comparison of some results of our pipeline with his pipeline and for offering helpful advice. ALK has been supported by NASA through Hubble Fellowship grant 51257.01 awarded by STScI, which is operated by AURA, Inc., for NASA under contract NAS 5-26555. We recognize and acknowledge the very significant cultural role and reverence that the summit of Mauna Kea has always had within the indigenous Hawaiian community. We are most fortunate to have the opportunity to conduct observations from this mountain.

\bibliographystyle{/users/akraus/Dropbox/Papers/apj.bst}

\bibliography{/users/akraus/Dropbox/Papers/krausbib.bib}

\clearpage

\begin{landscape}
\clearpage

\begin{deluxetable}{lcccl}
\tabletypesize{\scriptsize}
\tablewidth{0pt}
\tablecaption{Observing Log for LkCa 15}
\tablehead{
\colhead{Epoch (UT Date)} & \colhead{Filter} & \colhead{Visits\tablenotemark{a}} & \colhead{Seeing\tablenotemark{b}} & \colhead{Calibrators\tablenotemark{c}}
}
\startdata
2009 Nov 19&L'&6&0.9''&GM Aur (x10), HD 283572, HP Tau/G2 (x3), UX Tau (x6), V819 Tau\\
2009 Nov 20&L'&7&0.5''&DG Tau (x2), DR Tau (x2), HP Tau/G2 (x4), UX Tau (x8)\\
2010 Aug 16&L'&3&0.5''&GK Tau, HP Tau/G2\\
2010 Aug 17&L'&2&0.5''&DO Tau, HP Tau/G2\\
2010 Nov 26&K'&12&0.6''&CI Tau (x2), DO Tau (x2), DQ Tau, DS Tau, GK Tau (x2), HP Tau/G2 (x3), UX Tau (x2)\\
2010 Nov 27&L'&4&0.8''&DO Tau, DS Tau, GK Tau, HP Tau/G2, SU Aur, UX Tau\\
\enddata
\tablenotetext{a}{Each ``visit'' is a single observation consisting of 20 images (for $L'$) or 12 images (for $K'$), each of which has an integration time of 20s.}
\tablenotetext{b}{The seeing measurement we report is average value measured by the CFHT seeing monitor during the time of the observations, and is measured in the $V$ filter.}
\tablenotetext{c}{When calibrators were observed multiple times in one night, we denote this with a number in parentheses beside that calibrator's name.}
 \end{deluxetable}

\clearpage
 
 \begin{deluxetable}{lc|rrr|rrr|rrr|rr}
\tabletypesize{\tiny}
\tablewidth{0pt}
\tablecaption{Photometry and Astrometry}
\tablehead{
\colhead{Epoch} & \colhead{Filter} & 
\colhead{Sep (mas)} & \colhead{PA (deg)} & \colhead{$\Delta m$ (mag)} & 
\colhead{Sep (mas)} & \colhead{PA (deg)} & \colhead{$\Delta m$ (mag)} & 
\colhead{Sep (mas)} & \colhead{PA (deg)} & \colhead{$\Delta m$ (mag)} & 
\colhead{$\chi ^2$} & \colhead{$N_{df}$}
\\
\colhead{} & \colhead{} & \multicolumn{3}{c}{(NE)} & \multicolumn{3}{c}{(CEN)} & \multicolumn{3}{c}{(SW)}
}
\startdata
{\bf No Companions}&&&&&&&&&&&&\\
2009 November&L'&...&...&...&...&...&...&...&...&...&735&336\\
2010 August&L'&...&...&...&...&...&...&...&...&...&257&140\\
2010 November&L'&...&...&...&...&...&...&...&...&...&269&112\\
2010&L'&...&...&...&...&...&...&...&...&...&525&252\\
2009+2010&L'&...&...&...&...&...&...&...&...&...&1261&588\\
2010 November&K'&...&...&...&...&...&...&...&...&...&364&333\\

{\bf One Companion}&&&&&&&&&&&&\\
2009 November&L'&...&...&...&77.3$\pm$3.0&318.1$\pm$1.6&5.61$\pm$0.07&...&...&...&502&333\\
2010 August&L'&...&...&...&78.3$\pm$4.4&321.2$\pm$2.7&5.34$\pm$0.12&...&...&...&169&137\\
2010 November&L'&...&...&...&80.9$\pm$3.6&320.6$\pm$2.0&5.17$\pm$0.11&...&...&...&154&109\\
2010&L'&...&...&...&79.4$\pm$2.9&320.9$\pm$1.7&5.25$\pm$0.08&...&...&...&325&249\\
2009+2010&L'&...&...&...&78.0$\pm$2.1&318.7$\pm$1.2&5.49$\pm$0.05&...&...&...&837&585\\
2010 November&K'&...&...&...&65.6$\pm$1.5&332.1$\pm$1.2&6.67$\pm$0.09&...&...&...&364&333\\

{\bf Two Companions}&&&&&&&&&&&&\\
2009 November&L'&97.4$\pm$1.9&2.4$\pm$1.2&5.52$\pm$0.08&...&...&...&85.3$\pm$2.4&310.9$\pm$1.6&5.53$\pm$0.07&321&330\\
2010 August&L'&96.0$\pm$4.9&350.1$\pm$3.4&5.56$\pm$0.18&...&...&...&83.0$\pm$4.1&307.2$\pm$3.1&5.33$\pm$0.13&135&134\\
2010 November&L'&93.8$\pm$3.4&356.7$\pm$2.4&5.33$\pm$0.15&...&...&...&85.7$\pm$3.8&309.8$\pm$2.9&5.26$\pm$0.11&100&106\\
2010&L'&94.9$\pm$2.9&356.5$\pm$1.9&5.46$\pm$0.12&...&...&...&84.1$\pm$2.8&310.0$\pm$2.0&5.28$\pm$0.08&238&246\\
2009+2010&L'&96.5$\pm$1.6&1.0$\pm$1.1&5.51$\pm$0.07&...&...&...&84.6$\pm$1.9&311.0$\pm$1.2&5.45$\pm$0.06&571&582\\
2010 November&K'&...&...&...&67.5$\pm$1.6&333.3$\pm$1.3&6.68$\pm$0.09&88.9$\pm$2.6&302.9$\pm$1.4&7.20$\pm$0.13&296&330\\

{\bf Three Companions}&&&&&&&&&&&&\\
2009 November&L'&112.4$\pm$6.4&28.4$\pm$4.2&6.35$\pm$0.21&95.6$\pm$2.2&357.4$\pm$2.7&5.49$\pm$0.09&85.1$\pm$2.6&306.8$\pm$1.8&5.55$\pm$0.08&291&327\\
2010 August&L'&111.4$\pm$5.9&13.7$\pm$3.8&5.81$\pm$0.24&92.1$\pm$4.9&336.8$\pm$4.3&5.36$\pm$0.15&88.8$\pm$6.0&295.0$\pm$5.0&5.51$\pm$0.18&113&131\\
2010 November&L'&91.5$\pm$4.1&3.0$\pm$5.4&5.51$\pm$0.14&155.7$\pm$12.2&323.5$\pm$2.9&6.26$\pm$0.34&85.8$\pm$4.8&310.5$\pm$3.4&5.25$\pm$0.11&90&103\\
2010&L'&104.7$\pm$4.7&11.2$\pm$3.8&5.81$\pm$0.19&92.4$\pm$4.0&337.3$\pm$4.7&5.43$\pm$0.13&86.6$\pm$4.2&299.1$\pm$4.1&5.51$\pm$0.15&211&243\\
2009+2010&L'&106.8$\pm$3.2&15.0$\pm$2.7&5.89$\pm$0.14&92.8$\pm$2.5&344.7$\pm$3.6&5.65$\pm$0.09&87.4$\pm$2.2&302.6$\pm$2.1&5.56$\pm$0.08&519&579\\
2010 November&K'&67.0$\pm$3.2&12.3$\pm$2.8&7.40$\pm$0.19&64.4$\pm$1.5&334.8$\pm$1.5&6.59$\pm$0.09&82.5$\pm$2.4&302.3$\pm$1.5&7.06$\pm$0.12&265&327\\

{\bf Three (global)}&&&&&&&&&&&&\\
ALL&L'&100.7$\pm$1.9&5.6$\pm$1.1&5.69$\pm$0.08&71.9$\pm$1.6&335.2$\pm$1.3&6.30$\pm$0.19&88.2$\pm$1.8&304.7$\pm$1.2&5.73$\pm$0.09&851&912\\
ALL&K'&...&...&7.85$\pm$0.28&...&...&6.73$\pm$0.10&...&...&7.13$\pm$0.12&...&...\\
\enddata
\tablecomments{Point sources in each model fit as are designated as best corresponding to the northeast (NE), central (CEN), or southwest (SW) component of the system. However, some fits to individual epochs are not significant, causing some spurious detections at discrepant positions (denoted by photometric uncertainty of $>$0.2 mag or $SNR<5$). All subsequent analysis uses the properties in the global fit.}
 \end{deluxetable}

\clearpage

\end{landscape}

\end{document}